\documentclass[12pt]{article}
\usepackage{amsmath}
\usepackage{amssymb}
\usepackage{graphicx}
\newcommand{\GeV}      {~\mathrm{GeV}} 
 
\newcommand{\lang}{\left\langle} 
\newcommand{\rang}{\right\rangle} 
\newcommand{\sigmav}{\lang\sigma v\rang}
\newcommand{\gev}{\mathrm{GeV}}

\begin{document}
\begin{titlepage}
\vspace*{-2.5cm}
\rightline{LPT--Orsay 09/71}
\rightline{FTUAM 09-20}
\rightline{IFT-UAM/CSIC-09-42}   

\vspace{2cm}

\begin{center}

{\huge {\bf Antimatter signals of singlet scalar dark matter}}
\vspace{0.8cm}\\

{\large A. Goudelis$^{1,2}$,Y. Mambrini$^1$, C. Yaguna$^{3,4}$
}
\vspace{0.3cm}\\

$^1$ 
Laboratoire de Physique Th\'eorique, 
Universit\'e Paris-Sud, F-91405 Orsay, France
\vspace{0.3cm}\\

$^2$ 
Istituto Nazionale Fisica Nucleare, 
Sezione di Padova, I-35131, Padova, Italy
\vspace{0.3cm}\\

$^3$ 
Departamento de F\'isica Te\'orica C-XI,
Universidad Aut\'onoma de Madrid,\\

Cantoblanco, 28049 Madrid, Spain
\vspace{0.3cm}\\

$^4$ 
Instituto de F\'isica Te\'orica UAM/CSIC,
Universidad Aut\'onoma de Madrid,\\
Cantoblanco,
28049 Madrid, Spain
\vspace{0.3cm}\\

\end{center}

\vspace{0.5cm}

\begin{abstract}
We consider the singlet scalar model of dark matter and study the expected antiproton and positron signals from dark matter annihilations. The regions of the viable parameter space of the model that are excluded by present data are determined, as well as those regions that will be probed by the forthcoming experiment AMS-02. In all cases, different propagation models are investigated, and the possible enhancement due to dark matter substructures is analyzed. We find that the antiproton signal is more easily detectable than the positron one over the whole parameter space.  For a typical propagation model and without any boost factor, AMS-02 will be able to probe --via antiprotons-- the singlet model of dark matter up to masses of $600$ GeV. Antiprotons constitute, therefore, a promising signal to  constraint or detect the singlet scalar model.
\end{abstract}

\end{titlepage}

\newpage

\tableofcontents

\newpage

\section{Introduction}
One of the simplest extensions of the Standard Model that can explain the dark matter is the addition of a real scalar singlet and an unbroken $Z_2$ symmetry under which the singlet is odd while all other fields are even. Such a model, known as the singlet scalar model of dark matter, has been studied several times in the literature \cite{McDonald:1993ex, Burgess:2000yq, Davoudiasl:2004be, Barger:2007im, Dick:2008ah,Yaguna:2008hd}. Predictivity is undoubtedly its most salient feature. In contrast with other common scenarios that explain the dark matter, the singlet model contains only one additional field, the singlet scalar, and  two new parameters: the singlet mass and the coupling between the singlet and the  higgs boson --the only standard model field that couples to it. The singlet  relic density as well as its direct and indirect detections rates depend additionally on the higgs mass, which however cannot vary  in a wide range. After imposing the dark matter constraint, the viable parameter space gets reduced simply to the singlet mass and the higgs mass. The implications of the model can then be studied as a function of the singlet mass for a few representative values of the higgs mass. The singlet scalar model, therefore, provides a compelling and predictive scenario to explain  the dark matter.

Recently, it was shown \cite{Yaguna:2008hd} that the singlet may  have the right relic density to explain the observed dark matter abundance, that its direct detection cross section is large enough to be probed by present and planned experiments, and that the gamma rays from the annihilation of singlet scalar dark matter will likely be observed by  the Fermi satellite \cite{fermi}.  To complete the analysis of this model, in this paper we study the indirect detection of singlet scalar dark matter through positrons and antiprotons.

The indirect detection of dark matter is one of the promising avenues towards its identification. Among them, antimatter searches in cosmic rays play an essential role. Present experiments such as Pamela \cite{pamela} and Fermi \cite{fermi} are already measuring the cosmic ray spectrum with unprecedented precision. In fact they have already challenged our knowledge of the positron spectrum. Future experiments such as AMS-02 \cite{ams2}, which is scheduled to be launched next year, will measure the antiproton and the positron spectrum in a wider  energy range and with significantly better statistics. Hence, the chances of observing an exotic component from dark matter annihilation in the positron or the antiproton spectrum are now higher than ever. It is therefore crucial to study the expected antiproton and positron fluxes in well-motivated scenarios for physics beyond the standard model that account for the dark matter of the Universe.

The antiproton and positron fluxes from singlet annihilation not only depend on the dark matter model but also on a number of astrophysical factors that affect the production and propagation of such particles throughout the Galaxy. To take these effects into account, we consider different propagation models for positrons and antiprotons, and we study the possible role of substructures in the dark matter halo. Our goals are to find out whether any  region of the parameter space is already ruled out by present data from Pamela, and to identify  those regions that are within the sensitivity of the AMS-02 experiment. In addition, by comparing the corresponding regions for positrons and antiprotons, we expect to determine which of these two signals offers better perspectives to probe the singlet scalar model.

In the next section the main features of the  singlet scalar model of dark matter will be briefly reviewed. Then, in section \ref{sec:antip}, we present a detailed analysis of the antiproton signal in this model.  In particular, we determine the excluded and detectable regions in the plane ($m_S,\sigmav)$ for different sets of parameters accounting for propagation and substructure effects. In section \ref{sec:positrons} we present an analogous analysis for the positron channel. A brief discussion of our main results is given in section \ref{sec:disc}, which is followed by our conclusions. 

\section{The singlet scalar model of dark matter}
\label{sec:singlet}
The Lagrangian that describes the singlet scalar model of dark matter  is
\begin{equation}
\mathcal{L}= \mathcal{L}_{SM}+\frac 12 \partial_\mu S\partial^\mu S-\frac{m_0^2}{2}S^2-\frac{\lambda_S}{4}S^4-\lambda S^2 H^\dagger H\,,
\label{eq:la}
\end{equation}
where $\mathcal{L}_{SM}$ denotes the Standard Model Lagrangian, $H$ is the higgs doublet, and $S$ is the singlet scalar field. This  is the most general renormalizable Lagrangian that is compatible with  the $SU(3)\times SU(2)\times U(1)$ gauge invariance and with the symmetry $S\to -S$. Notice, from the Lagrangian, that the higgs boson is the only standard model field that directly couples to  the singlet. This  extension of the standard  model contains, therefore,   two  new phenomenologically relevant parameters: $m_0$ and $\lambda$. Instead of  $m_0$, it is useful to consider the physical mass of the singlet
\begin{equation}
m_S=\sqrt{m_0^2+\lambda v_{EW}^2}
\end{equation}
as a free parameter of the model. The other free parameter is  $\lambda$, which determines the  trilinear ($S^2h$) and quartic ($S^2h^2$) coupling between the singlet and the physical higgs boson. In addition to $m_S$ and $\lambda$, the dependence on the higgs mass --a SM parameter-- should also be also taken into account, as it affects  the dark matter phenomenology.  To accurately compute the singlet relic density we use the micrOMEGAs package \cite{Belanger:2006is}, which can calculate the relic density in a generic dark matter model.

\begin{figure}[tb]
\centering
\includegraphics[scale=0.5]{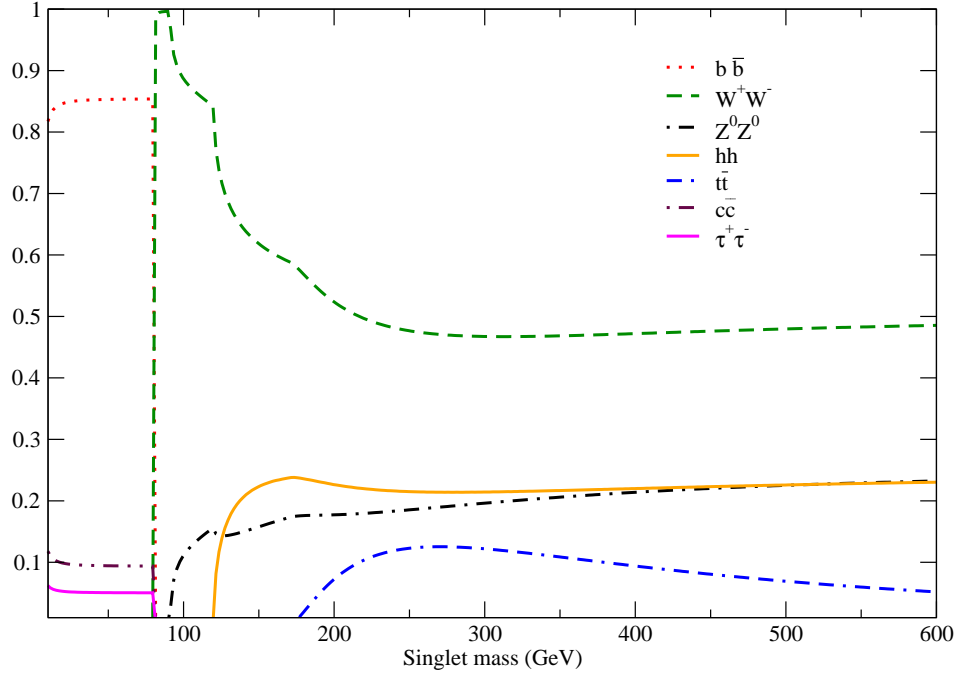}
\caption{ Annihilation branching fractions as a function of the dark matter mass for the singlet model. The higgs mass was set to $120\GeV$ while the value of $\lambda$ was obtained by imposing the dark matter constraint. \label{fig:brs} }
\end{figure}

Singlets can annihilate through s-channel higgs boson exchange into a variety of final states: $f\bar f$, $W^+W^-$, $Z^0Z^0$, and $hh$. Additionally, they can also annihilate into $hh$ either directly or through singlet exchange. Depending on the singlet mass, two annihilation regions can be clearly distinguished. A light singlet, $m_S<m_W$, will annihilate mainly into the $b\bar b$ final state, with subdominant contributions from other light fermions. To obtain the correct relic density, such light singlets require a relatively large value of $\lambda$ ($\gtrsim 0.1$), and are consequently constrained by direct detection experiments. Singlets masses below $50\GeV$, for instance, are already ruled out by current measurements \cite{Yaguna:2008hd}. A heavy singlet, $m_S>m_W$, on the other hand, will annihilate mainly into $W^+W^-$, with additional contributions from $Z^0Z^0$, $hh$ and $t\bar t$. In this region the required value of $\lambda$ is typically smaller ($\sim 0.01$) and present direct detection constrains are ineffective. Figure \ref{fig:brs} displays the annihilation branching fractions as a function of the singlet mass for $m_h= 120\GeV$. The sharp contrast between the light singlet region and the heavy one is clearly observed. 

Being a scalar field, the singlet annihilation rate today is not suppressed with respect to that of the early Universe (s-wave annihilation). Moreover, throughout most of the viable parameter space $\langle\sigma v\rangle$ is constant and equal to the typical annihilation rate, $\langle\sigma v\rangle\sim 2.3\times 10^{-26}\mathrm{cm}^3\mathrm{s}^{-1}$ \cite{Yaguna:2008hd}. The only two places where $\langle\sigma v\rangle$ is smaller are at the $W$ threshold ($m_S=m_W-\epsilon$) and at the higgs resonance ($m_h\sim 2m_S$). In fact, the main effect of the higgs mass is to determine the position where this resonance lies. In the following analysis we will set $m_h$ to $120\GeV$ and consider only the remaining dependence on the singlet mass. For the sake of definiteness,  singlet masses below $600\GeV$ will be considered.
\section{Antiprotons}
\label{sec:antip}
Antiprotons can be produced either by dark matter annihilations 
or through different kinds of astrophysical mechanisms. After being produced they propagate in the Galaxy and reach the Earth, where they can be detected as exotic components in cosmic rays. The PAMELA experiment reported recently \cite{Adriani:2008zq} the  measurement of  the antiproton to proton flux ratio up to $100\GeV$, which is in agreement with that predicted by the  conventional background model. In the near future, the AMS-02 experiment, to be launched next year, will measure the antiproton flux with even higher precision, increasing the odds of finding an additional component due to dark matter annihilations. 

\subsection{Propagation}
Once produced, antiprotons propagate throughout the galaxy undergoing several processes:

\begin{itemize}
\item They scatter on irregularities of the galactic magnetic field --Alfv\'en waves.
These scatterings constitute, in fact, a space diffusion process with a diffusion coefficient given by
\begin{equation}
K(E_{\mbox{\tiny{kin}}}) = K_0 \beta_{\bar{p}}
\left( \frac{p_{\bar{p}}}{\mbox{GeV}} \right)^\delta\,,
\end{equation}
where $E_{\mbox{\tiny{kin}}}$ is the antiproton kinetic energy, 
$p_{\bar{p}}=(E_{\mbox{\tiny{kin}}}^2 + 2m_{\bar{p}}E_{\mbox{\tiny{kin}}})^{1/2}$ 
is the antiproton momentum, and 
$\beta_{\bar{p}} = \left(  1 - \frac{m_{\bar{p}}^2}{(E_{\mbox{\tiny{kin}}} + m_{{\bar{p}}})^2} \right)^{1/2}$. $K_0$ 
and $\delta$  are free parameters of the propagation model that  are constrained by  a combination of 
theoretical predictions and astrophysical data.

\item They undergo a second order Fermi mechanism reacceleration, due to the motion of the scattering centers with a velocity of
$V_a \approx (20 - 100)$ km/sec. This reacceleration process is described by the coefficient
\begin{equation}
K_{EE} = \frac{2}{9} V_{a}^{2} \frac{E^2 \beta ^4}{K(E)}\,.
\end{equation}

\item They lose energy either adiabatically, or through Coulomb scattering, or by
ionizing the interstellar (IS) medium. The total energy loss rate is denoted by $b$ and depends on the antiproton energy.

\item They are wiped away from the galactic disk through convection, with a 
velocity $V_c \approx (5 - 15)$ km/sec. In the following, this velocity will 
be taken to be completely vertical to the disk: $V_c = V_c \mbox{sign}(z)$.

\item They can annihilate upon scattering on the IS medium. In this study, 
we shall consider the two primary components of the medium, namely Hydrogen
and Helium. The annihilation cross-sections for $\bar{p}-\mbox{H}$ and 
$\bar{p}-\mbox{He}$ scattering
have been taken from \cite{CrossSections}, where the well-known Tan\&Ng
parametrization \cite{TanNG} is used.
\end{itemize}
The propagation of antiprotons in the interstellar medium, taking into account all of the above processes, is described by a diffusion-convection equation of the form
\begin{equation}
\partial_z(V_c\psi) - K\nabla\psi + \partial_E \left[ b(E)\psi - K_{EE}(E)\partial_E \psi \right] = q\,,
\label{masterProp}
\end{equation}
where we denote by $\psi = dn/dE$ the energy density of the antiprotons, and by  $q$  the source term --see equation (\ref{eq:q}). Actually, this equation is rather generic and can be applied to other particles species propagating in the galaxy. Only the values of the different parameters, but not the equation itself, will vary depending on the propagating particle. We will see in  section \ref{sec:positrons}, for instance, that this same equation describes the propagation of positrons in the interstellar medium. Let us now see how to solve this equation for antiprotons.

\subsection{The primary flux}
To solve equation (\ref{masterProp}), we shall use the method proposed, 
for example, in
\cite{FullCalcClumps,GalacticLottery,BringmannSalati}. The idea of this method is to adopt a simpler version of equation (\ref{masterProp})
for which the Green function can be calculated analytically. To achieve  such simplification, certain processes contributing to the final antiproton spectrum have to be  neglected. Specifically,
all energy redistributions in the initial (injection) spectrum --energy losses, reacceleration, as well as "tertiary" contributions (i.e. contributions
from secondary antiprotons produced upon inelastic scattering with the IS medium)-- are ignored. Whether these redistributions are important or not depends mainly on the antiproton energy. For GeV energies, the results may deviate up to $50\%$ from those obtained with the  (more complete) Bessel function treatment --in \cite{FastFormulae} a comparison between the two methods
can be found (see figure 2). But for energies around $~10$ GeV, the accuracy
of the method improves dramatically, yielding essentially indistinguishable
 results at slightly higher energies. Since the $\bar{p}$ energy region we shall consider
begins at $10$ GeV, we can safely use this simplified approach. Apart from its clearer physical interpretation, the main advantage of this method is that astrophysical effects can be separated from particle physics ones. As a result, it is a well-suited method for scans in parameter spaces, as those we are going to carry out in the following.

If we denote by $\Gamma_{\overline{p}}^{\mbox{\tiny{ann}}} = \sum_{\mbox{\tiny{ISM}}} 
n_{\mbox{\tiny{ISM}}} v \sigma_{\overline{p} \ \mbox{\tiny{ISM}}}^{\mbox{\tiny{ann}}}$
the destruction rate of antiprotons in the ISM, where $\mbox{ISM} = \mbox{H, He}$, 
and implementing the aforementioned simplifications, 
the transport 
equation for a point source (which actually defines the propagator $G$) is:
\begin{equation}
\left[ -K\nabla + V_c\frac{\partial}{\partial z}
+2 h \Gamma_{\mbox{tot}} \delta(z) \right] G = 
\delta(\vec{r} - \vec{r}')\,.
\end{equation}
The antiproton propagator can then be written as
\begin{equation}
G^{\odot}_{\overline{p}}(r,z) = 
\frac{e^{-k_v z}}{2 \pi K L}
\sum_{n=0}^{\infty} c_n^{-1} K_0(r\sqrt{k_n^2 + k_v^2})
\sin(k_n L) \sin(k_n(L-z))\,,
\label{GreenPbars}
\end{equation}
where
$K_0$ is a modified Bessel function of the second kind and
\begin{eqnarray}{\nonumber}
c_n & = & 1 - \frac{\sin(k_n L) \cos(k_n L)}{k_n L}\,,\\
k_v & = & V_c/(2K)\,,\\ \nonumber
k_d & = & 2h \Gamma_{\overline{p}}^{\mbox{\tiny{ann}}}/K + 2 k_v\,, \nonumber
\end{eqnarray}
with $h = 100$ pc being the half thickness of the galactic disc, and
$L$ being the half-thickness of the diffusive zone.  $k_n$ is obtained as the solution of the equation
\begin{equation}
n\pi - k_{n}L - \arctan(2k_n/k_d) = 0, \ \ n\in\mathbb{N}\,. \nonumber
\end{equation}
Then, in order to compute the flux expected on  earth, we should
convolute the Green function (\ref{GreenPbars}) with the source 
distribution $q(\vec{r}, E)$. For dark matter annihilations in the galactic halo, the source term is given by
\begin{equation}
q(\vec{r}, E) = \frac{1}{2}
\left( \frac{\rho(\vec{x})}{m_\chi} \right)^2
\sum_i 
\left( 
\langle\sigma v\rangle \frac{dN_{\bar{p}}^i}{dE_{\bar{p}}}
\right)\,,
\label{eq:q}
\end{equation}
where the index $i$ runs
over all possible annihilation final states. The decay of SM 
particles into antiprotons can be calculated with programs like
PYTHIA  \cite{PYTHIA}. In the singlet scalar model of dark matter, antiprotons may originate in $W^\pm$, $Z^0$, $h$, and $t$ decays. Regarding the distribution of dark matter in the Galaxy, $\rho(\vec{x})$, we assume a NFW profile with a local density of $0.3\GeV \mathrm{cm}^3$. The final expression for the antiproton flux on the Earth takes the form
\begin{equation}
\Phi_{\odot}^{\bar{p}} (E_{\mbox{\tiny{kin}}}) = 
\frac{c\beta }{4\pi}
\frac{\langle\sigma v\rangle}{2}
\left(   \frac{\rho(\vec{x}_{\odot})}{m_\chi} \right)^2
\frac{dN}{dE}(E_{\mbox{\tiny{kin}}})
\int_{DZ} \left(\frac{\rho(\vec{x_s})}{\rho(\vec{x}_{\odot})} \right)^2
G^{\odot}_{\overline{p}}(\vec{x}_s) d^3x\,,
\label{PbarFlux}
\end{equation}
where none of the integrated quantities depends on the antiproton energy. This  feature  demonstrates one of the virtues of the Green 
function method applied to antiprotons: The integral in equation (\ref{PbarFlux}), which we compute using a VEGAS Monte-Carlo algorithm, needs to be calculated only once for each value of the injection energy, for it is the same as the detection energy.

Regarding  the propagation parameters $L, K_0, \delta$, and  $V_c$, we take their values from 
the well-established MIN, MAX and MED models --see table \ref{PropParameters}. The former two models correspond to the minimal and maximal antiprotons fluxes that are compatible with the B/C data. The MED model, on the other hand, correspond to  the parameters that best fit the B/C data.
 
\begin{center}
\begin{table}
\centering
\begin{tabular}{|c|cccc|}
\hline 
&$L$ (kpc)&$K_0$(kpc$^2$/Myr)&$\delta$&$V_c$(km/s)\\
\hline 
MIN & $1$ & $0.0016$ & $0.85$ & $13.5$\\ 
MED & $4$ & $0.0112$ & $0.70$ & $12.0$\\
MAX & $15$ & $0.0765$ & $0.46$ & $5.0$\\
\hline 
\end{tabular}
\caption{{\footnotesize Values of propagation parameters
widely used in the literature and providing minimal and maximal antiproton fluxes,
or consitute the best fit to the B/C data.}}
\label{PropParameters}
\end{table}
\end{center}

\subsection{Influence of substructures}
N-body simulations reveal that galactic halos are not completely smooth; they also  contain a significant  number of substructures (clumps).  Such substructures have been studied repeatedly in the literature as a possible way to  enhance the dark matter annihilation rate. It has been claimed though, that in realistic scenarios it is rather improbable to expect large enhancements from dark matter clumps \cite{FullCalcClumps}. Here, we will study rather qualitatively the possible effect of substructures on the antiproton (and positron) signal.

To that end, we closely follow the approach outlined in \cite{GalacticLottery}. The enhancement due to dark matter substructure is described by an energy-dependent function known as the boost factor $B$. Because the distribution of dark matter clumps in the Galaxy  is unknown, $B$ cannot be computed from first principles; it can only be studied from a statistical point of view. For simplicity, we will limit our discussion to  the effective boost factor, $B_{eff}$, defined as the average value of $B$ over a large number of realizations of the Galactic halo. It must be kept in mind, however, that in some exceptional cases --for example when there is a large dark matter clump very close to the Earth-- $B_{eff}$ can deviate significantly from $B$. Since  the probability of such an event is quite small \cite{Hooper:2003ad}, we will not consider those cases in this work. 

The energy-dependent effective boost factor, $B_{eff}$, can be written, under certain assumptions (see  \cite{GalacticLottery}), as
\begin{equation}
B_{\mbox{\tiny{eff}}} \equiv \frac{\langle\phi\rangle}{\phi_{\mbox{\tiny{sm}}}} = 
(1-f)^2 + f B_c \frac{{\cal{I}}_1}{{\cal{I}}_2}\,,
\label{EffBoost}
\end{equation}
where $\langle\phi\rangle$ is the average flux coming from the clumpy DM distribution, $\phi_{\mbox{\tiny{sm}}}$ is the 
flux that we would expect if the whole halo were smooth, $f$ is the
fraction of DM in clumps, and  $B_c$ is the boost factor (assumed constant with energy)
for an individual clump.  In this study, this constant boost factor is supposed to be universal for all clumps. Finally, ${\cal{I}}_{n=1,2}$ are given by 
\begin{equation}
{\cal{I}}_n = 
\int_{\mbox{\tiny{DM halo}}} 
G(\vec{x},E) \left( \frac{\rho_{\mbox{\tiny{sm}}}(\vec{x})}{\rho_0} \right)^n d^3\vec{x}\,.
\end{equation}
The effective boost factor, then, depends on $f$ and $B_c$. When invoking  clumpiness, we will follow \cite{GalacticLottery} and use $f=0.2$ as a representative value (see e.g. \cite{Bertone:2005xz,Diemand:2005vz}). Regarding the constant boost factor, $B_c$, it could vary from just a few up to two orders of magnitude \cite{FullCalcClumps,Berezinsky:2003vn,Diemand:2005vz}. We will use  $B_c=3, 10, 100$, which give rise to effective boost factors in the approximate ranges ($1,2$), ($3,5$) and ($10,40$) respectively. This last range roughly coincides with the upper limit for the boost factor found in \cite{FullCalcClumps}, for the case of a NFW smooth halo and clumps with a Moore et al internal profile. 


\subsection{Astrophysical backgrounds}
So far, our main concern has been the antiproton signal from dark matter annihilations. But, to determine whether a model is detectable or whether it is compatible with present data, we also need to know the antiproton background. The most well-known treatment of the astrophysical antiproton background is that of Strong and Moskalenko \cite{CrossSections}, which has recently received additional experimental support. In fact, the antiproton data \cite{Adriani:2008zq} from the PAMELA experiment is perfectly  compatible with their predictions. We thus have a relatively simple antiproton background model that is able to explain the data.

In the following, when trying to examine whether the model is excluded by the PAMELA data, we shall be employing the theoretical background predictions, normalizing them to the low-energy PAMELA data. More specifically, we take advantage of the fact that predictions demonstrate a relatively constant background with respect to the propagation
model, and use the simple parametrization (properly normalized) presented in \cite{FastFormulae}. When studying the detectability of the model in AMS-02, on the other hand, we shall use the PAMELA measurements themselves as  background. More concretely, we borrow the fit performed by Cirelli et al in \cite{MinimalDM}, which provides a sufficiently good fit to theoretical predictions but also to 
the recent PAMELA data. We pay special attention at reproducing the good 
background normalization at low energies, so as to stay as close as possible to the PAMELA measurements.

Finally, let us point out that since the antiproton background does not vary significantly with the propagation parameters \cite{FastFormulae}, it is reasonable to consider a global background, irrespective of the propagation model. We will do so in the following.

\subsection{Results}
In this section we present our results for antiprotons. First, we combine the predicted flux and the expected background to determine the regions in the plane ($m_S, \sigmav$) that are already excluded by present antiproton data from PAMELA. Then, we determine the regions that are detectable by AMS-02. 

In the singlet scalar model of dark matter, the singlet mass and the higgs mass are the only free parameters left after imposing the relic density constraint. Since we take $m_h=120\GeV$ throughout this work, the annihilation branching ratios and the total annihilation rate, $\sigmav_{singlet}$, depend only on $m_S$. To obtain the excluded  regions in the plane ($m_S,\sigmav$), we first compute, for a model with the same branching ratios as the singlet scalar model (see figure \ref{fig:brs}), 
the value of $\sigmav$ required to exclude the model, $\sigmav_{excl}$.  By comparing $\sigmav_{singlet}$ with $\sigmav_{excl}$ we can then  determine whether the model is excluded or not at a given singlet mass. An analogous procedure is followed to determine the detectable regions.

\begin{figure}[tb!]
\centering
      \includegraphics[width=0.40\textwidth,angle=-90]{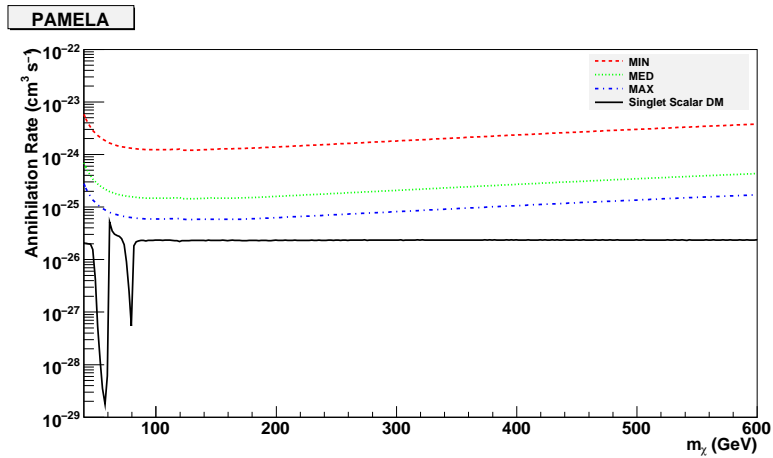}
      \caption{\footnotesize Regions of the parameter space that are  excluded by the  recent antiproton data from the PAMELA experiment. The area above the MIN, MED, and MAX lines is excluded for the given propagation model. The solid (black) line shows the viable parameter space of the singlet scalar model of dark matter. }
       \label{fig:pamela}
\end{figure}

The condition we impose for exclusion is that the sum of the signal prediction and the (properly normalized) background prediction exceed the measurement by the PAMELA collaboration by more than $3\sigma$. Figure \ref{fig:pamela} shows the excluded region for the MIN, MED, and MAX propagation models. The area above the lines is already excluded by present data. Also shown as a solid (black) line is the region along which the relic density constraint is satisfied --the viable parameter space of the model.  Because that line is entirely below the exclusion lines, we conclude that, in the absence of substructure enhancements,  current antiproton data do not yet constrain the singlet scalar model of dark matter. This situation may change soon, however, for the MAX  exclusion line lies just above the prediction of the model for masses around $100$ GeV. Future data from PAMELA may start constraining that region of the parameter space.

To assess the sensitivity of AMS-02 to an antiproton signal from singlet dark matter annihilations, we compute the annihilation rate needed to distinguish the signal from the background at the $95\%$ confidence level. To do so, we first calculate the $\chi^2$ as
\begin{equation}
\chi^2= 
\sum_{n=0}^N
\frac{(\phi^{tot}_n-\phi^{bkg}_n)^2}{(\phi^{tot}_n)}A\cdot T\ ,
\label{chi2}
\end{equation}
where $\phi^{tot}$ is the total antiproton flux, $\phi^{bkg}$ is the 
background flux, $N$ is the number of energy bins considered, $A$ is the
geometrical acceptance of the experiment, and $T$ is the data acquisition time. AMS-02 is expected to take data for  three years and features an antiproton  geometrical acceptance of $330$ cm$^2$sr \cite{Goy:2006pw}. We consider $20$ energy bins evenly distributed in logarithmic scale between $10$ and $300$ GeV. A $95\%$ confidence level corresponds then to  $\chi^2>31$.

\begin{figure}[tb!]
\centering
      \includegraphics[width=0.40\textwidth,angle=-90]{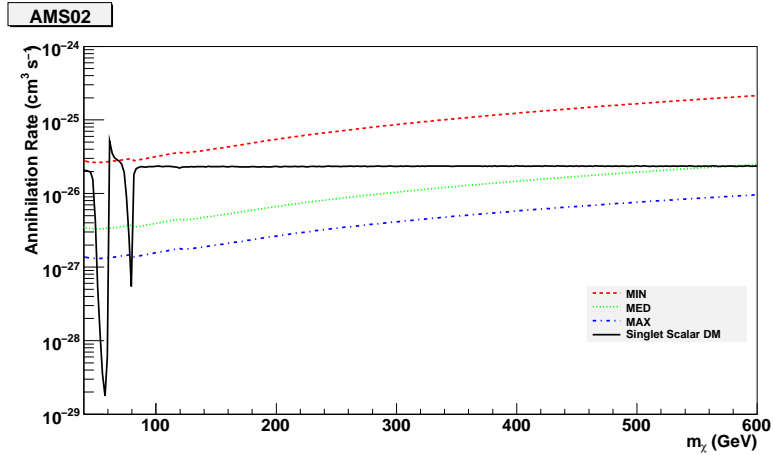}
      \caption{\footnotesize Regions of the parameter space that are within the sensitivity of the AMS-2 experiment. The area above the MIN, MED, and MAX lines is detectable by AMS-02. The solid (black) line shows the prediction of the singlet model. Notice that for MED and MAX, essentially the whole parameter space is detectable. }
       \label{fig:ams}
\end{figure}

The annihilation rates needed for a detection at AMS-02 are shown in  figure \ref{fig:ams}
for the three propagation models. The singlet model is detectable whenever its annihilation cross-section (solid line) is larger than the exclusion lines at AMS-02 for a given propagation model. We see that for the MIN propagation model only a small region around $60$ GeV  is within the AMS-02 reach. For the MED and MAX propagation models, on the other hand, essentially the whole parameter space is detectable at AMS-02. Only in the higgs resonance region, $2m_S\sim m_h$, and at the $W$-threshold does the model's annihilation rate fall below the exclusion lines. Hence, for the MED and MAX models, a singlet scalar with a mass below $600$ GeV should be easily detected at AMS-02.

\begin{figure}[tbp!]
\centering
	\includegraphics[width=0.40\textwidth,clip=true,angle=-90]{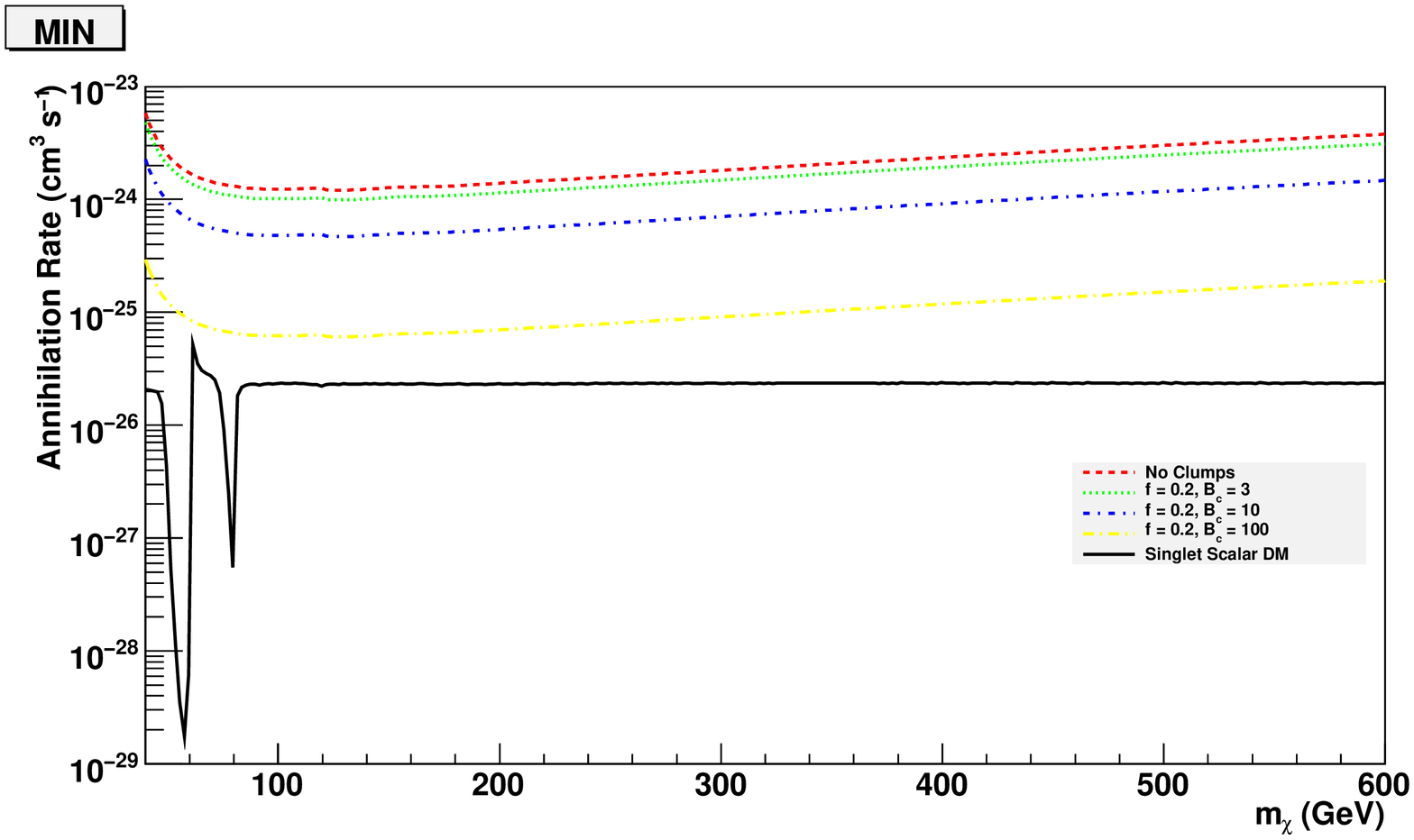}
	\includegraphics[width=0.40\textwidth,clip=true,angle=-90]{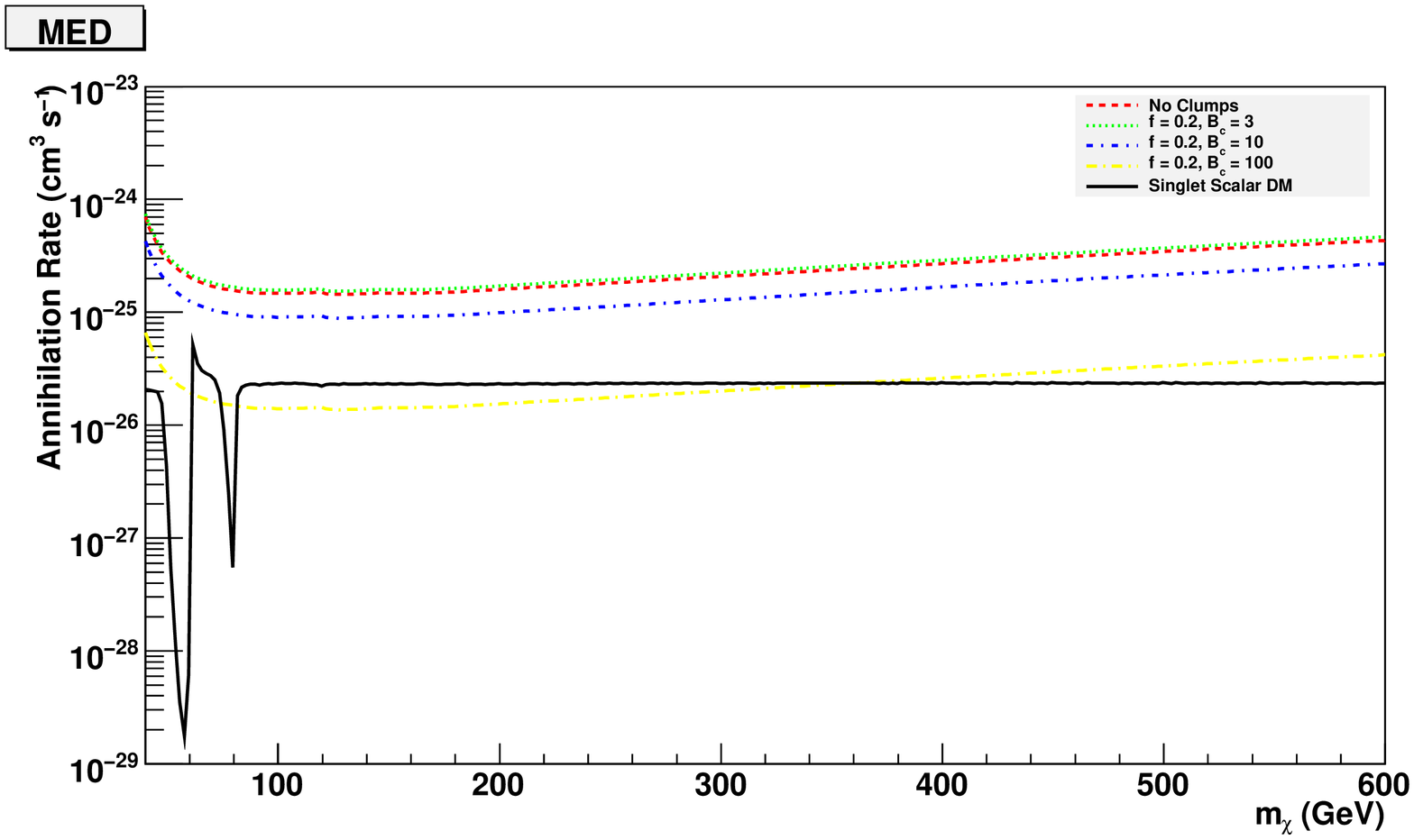}
	\includegraphics[width=0.40\textwidth,clip=true,angle=-90]{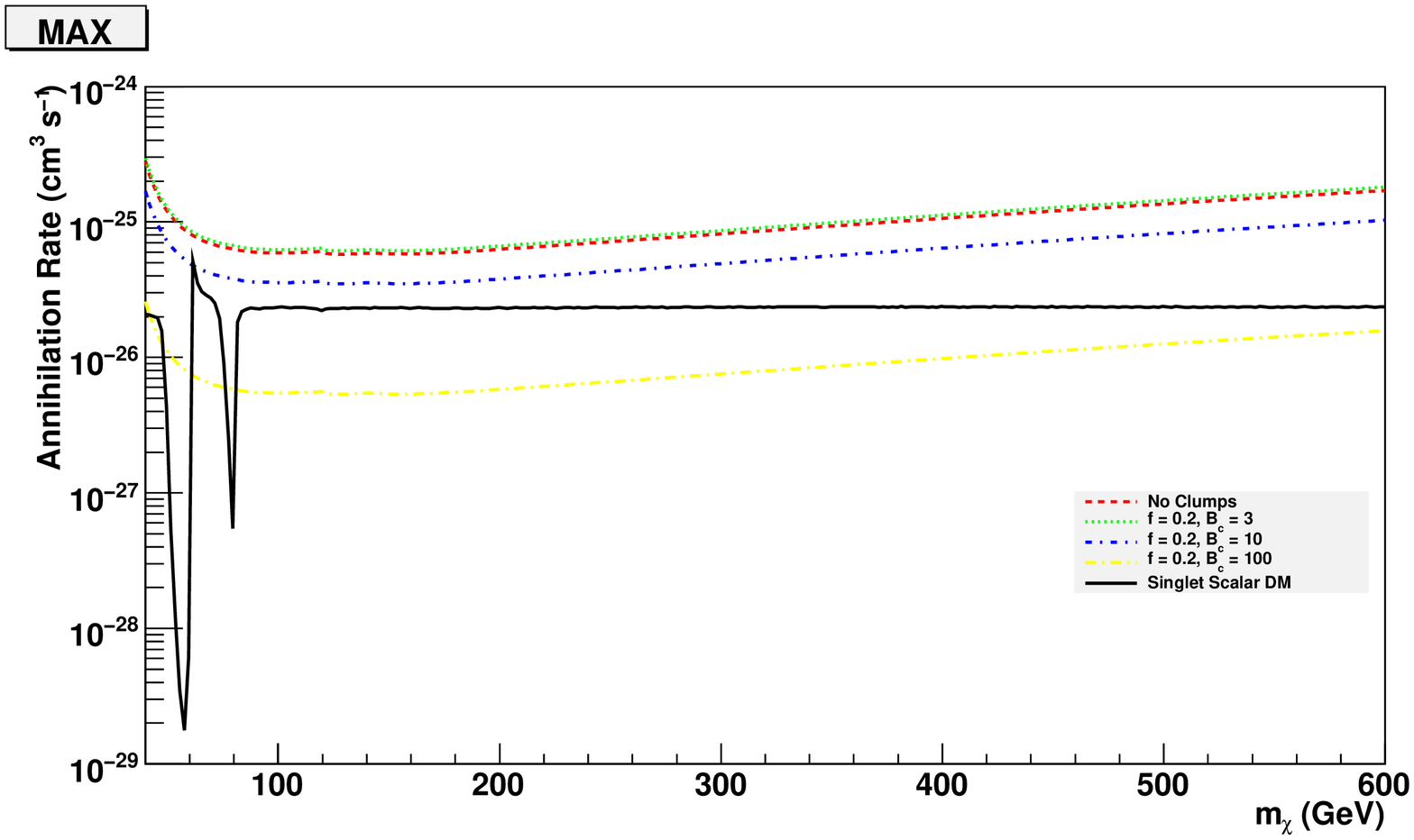}

          \caption{{\footnotesize
Regions excluded by the antiproton data from PAMELA including the possible effect of substructures in the DM halo. From top to bottom the figures correspond to the MIN, MED, and MAX propagation models. The solid (black) line shows the prediction of the singlet model. The area above the lines is excluded for the corresponding parameter values.}}
         \label{fig:ClumpsPbarsPAMELA}
\end{figure}

We now consider the possible effect of dark matter substructures on the exclusion and the detectable regions.  Figure \ref{fig:ClumpsPbarsPAMELA} shows the exclusion lines for the MIN (upper figure), MED (middle figure), and MAX (lower figure) propagation models for the different values of $B_c$ we investigate. As before, the solid (black) line shows the prediction of the singlet model. Notice that  the MIN propagation model is nowhere excluded,  not even for the most optimistic enhancement. For the MED propagation model, $B_c=100$ is excluded for singlet masses below $360$ GeV. The other values of $B_c$ are all compatible with present data. For the MAX propagation model, $B_c=100$ is ruled out in the whole mass range while $B_c=10$ is at the limits of exclusion for masses around $100$ GeV.  Hence, we see that, once the possible enhancement due to dark matter substructures is included,  current antiproton data already constrain some combinations of astrophysical  and particle physics parameters.    

\begin{figure}[tbp!]
\centering
	\includegraphics[width=0.40\textwidth,clip=true,angle=-90]{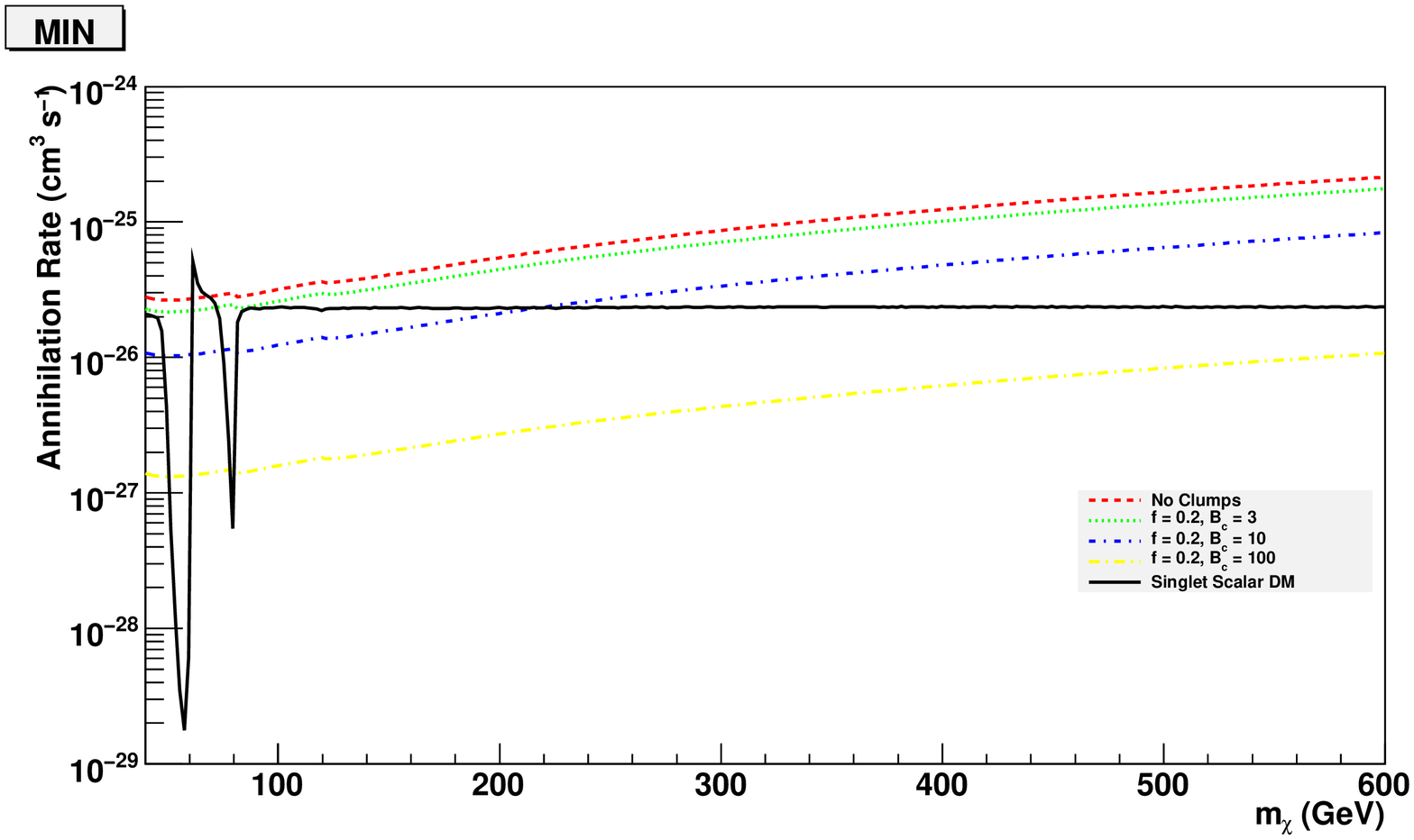}
	\includegraphics[width=0.40\textwidth,clip=true,angle=-90]{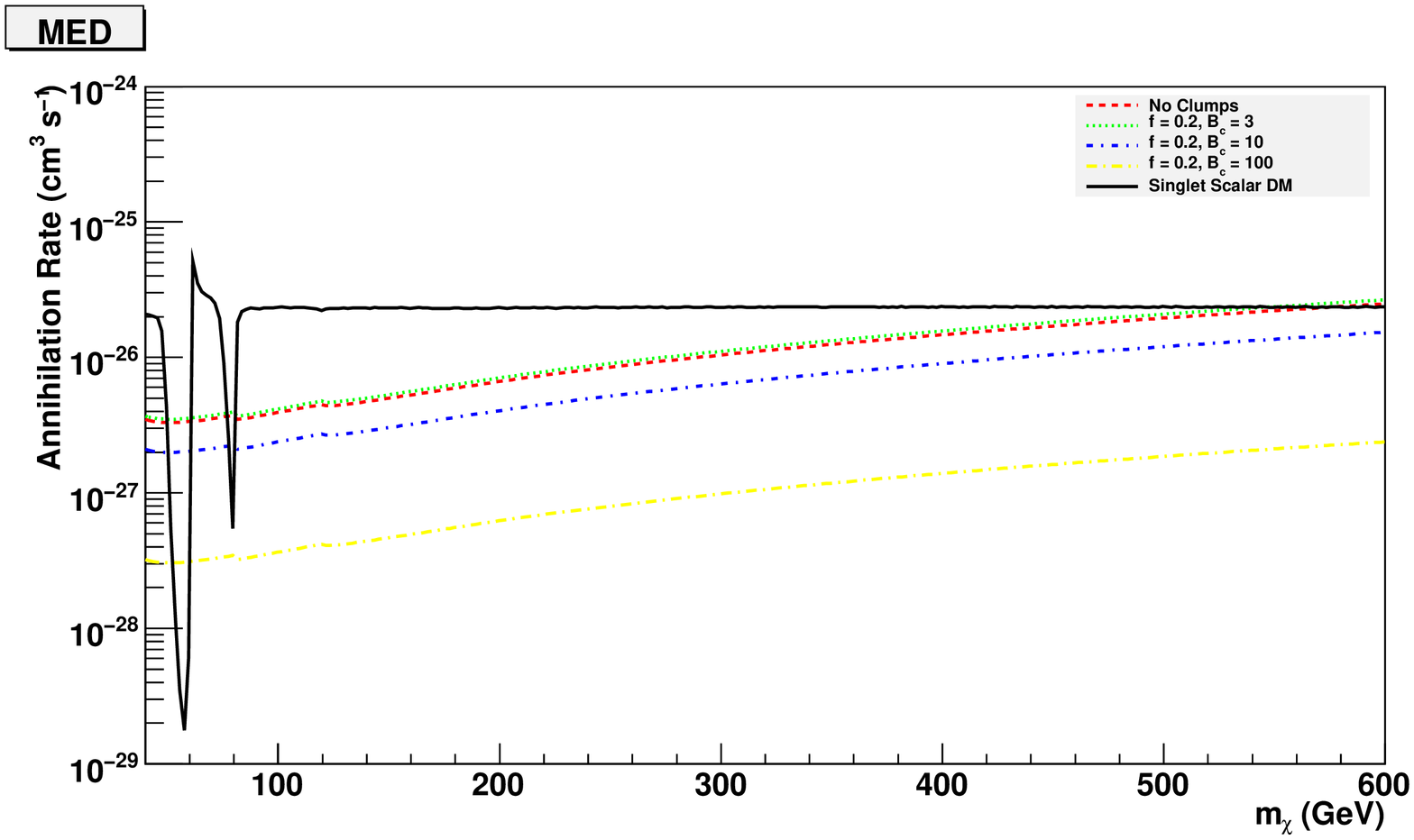}
	\includegraphics[width=0.40\textwidth,clip=true,angle=-90]{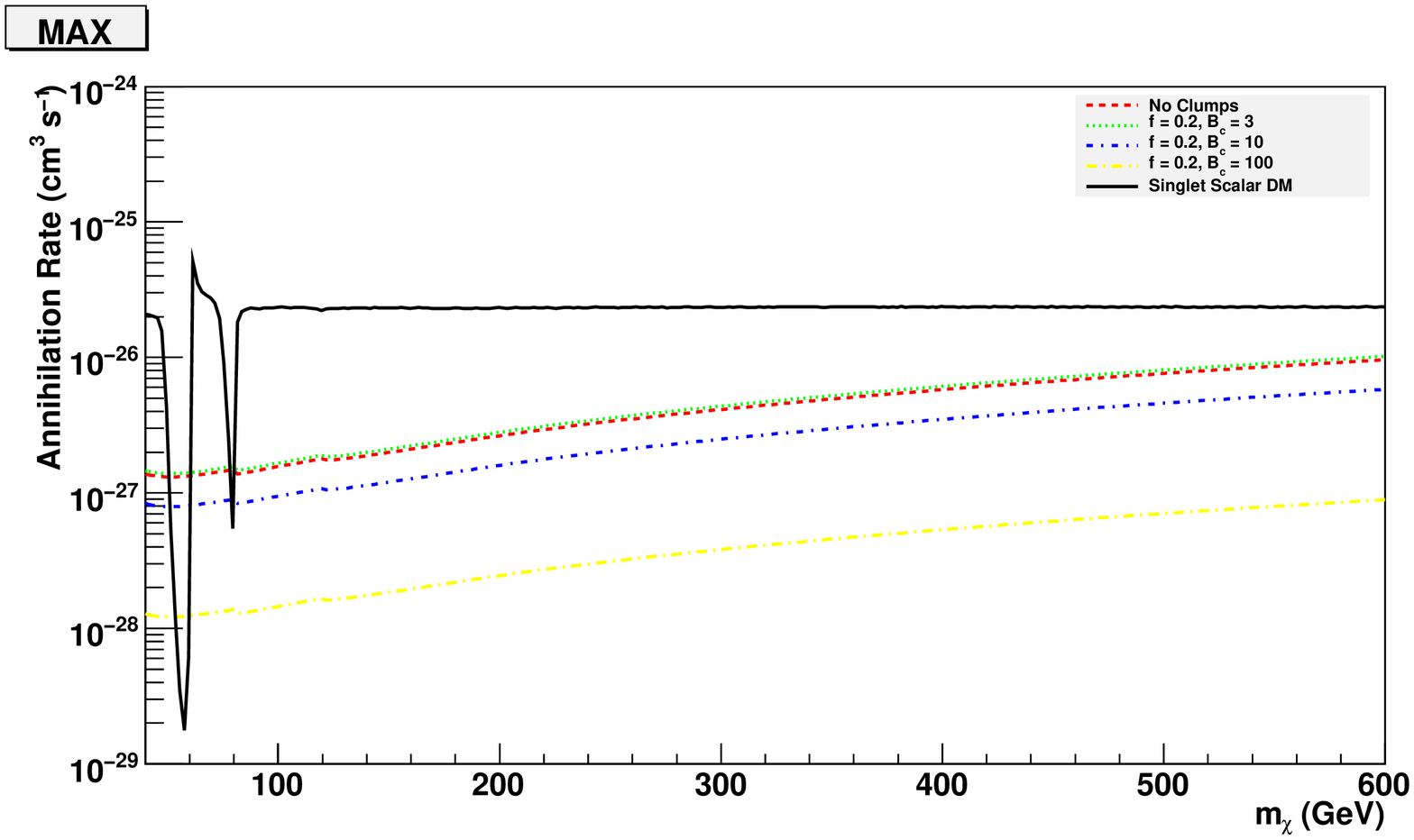}

          \caption{{\footnotesize
Detectable regions at AMS-02 including the possible effect of substructures. From top to bottom the figures correspond to the MIN, MED, and MAX propagation models. The solid (black) line shows the prediction of the singlet model. The area above the lines is detectable for the corresponding parameter values.  }}
        \label{fig:ClumpsPbarsAMS}
\end{figure}

The prospects for the detection of an antiproton signal at AMS-02 were already good without any additional boost from dark matter clumps, as we saw in figure \ref{fig:ams}. The additional enhancement from dark matter substructures only improves  such prospects, as clearly seen in figure \ref{fig:ClumpsPbarsAMS}. For the MIN model, for instance, the detectable region increases from a small region around $m_S\sim 60$ GeV without boost factor to $m_S\lesssim 200$ GeV for a moderate boost factor, covering the whole mass range for $B_c=100$.

To summarize, we have seen that the antiproton signal from singlet annihilation is a promising way to indirectly detect singlet scalar dark matter. Current antiproton data from PAMELA already constraint some scenarios and future data from AMS-02 is likely to reveal a significant excess over the expected background. 

\section{Positrons}
\label{sec:positrons}

We now discuss the positron signal from singlet dark matter annihilation, carrying out an analysis similar to that presented for antiprotons in the previous section. 
\subsection{Propagation}
There have been many treatments on the propagation of positrons throughout the galactic medium. In this paper, we adopt the 
two-zone diffusion model and solution described in \cite{GalacticLottery}. For completeness, 
we review here the main points of this approach.

Being a diffusive process, positron propagation is governed by the same general equation, (\ref{masterProp}), that describes antiproton propagation. The approximations and the parameters that were used for antiprotons, however,
are not the same as for positrons. Indeed, contrary to the $\overline{p}$ case, where energy redistribution processes 
become inefficient above a few GeV, energy loss --through either synchrotron radiation or inverse Compton 
scattering on stellar light and CMB photons-- is the main process 
involved in positron propagation. These processes lead to an energy loss rate that can be written as 
\begin{equation}
b(E) = \frac{E^2}{E_0 \tau_E}\,,
\end{equation}
where $E$ is the positron energy, $E_0$ is a reference energy (which we take to be 1 GeV) and $\tau_E=10^{16}$s is the characteristic energy-loss time.

The other important difference with respect to antiprotons is that for positron propagation the effect of the galactic convective wind,
as well as reacceleration and annihilation processes can all be neglected. After these simplifications are taken into account, we are left with the following equation
\begin{equation}
\partial_t \psi - \nabla \left[ K(x,E) \nabla\psi \right]- \partial_E \left[b(E)\psi\right] = q(x,E)\,,
\end{equation}
where $K$ is the space diffusion coefficient --steady state is assumed. This coefficient is  taken to be constant in space but depends on the energy as
\begin{equation}
K(E) = K_0\left(  \frac{E}{E_0}\right) ^\alpha.
\end{equation}
Here the diffusion constant, $K_0$, and the spectral index, $\alpha$, are propagation parameters.
Then, the master equation for positron propagation gets simplified to its final form
\begin{equation}
K_0 \epsilon^\alpha \bigtriangleup \psi  + 
\frac{\partial}{\partial \epsilon}\left( \frac{\epsilon^2}{\tau_E} \psi \right) + q = 0,
\label{masterPos}
\end{equation}
where $\epsilon=E/E_0$. This is the equation we  solve  to calculate the effects of positron propagation on a signal produced 
at some point in the galaxy.

\begin{center}
\begin{table}
\centering
\begin{tabular}{|c|ccc|}
\hline 
&$L$ (kpc)&$K_0$(kpc$^2$/Myr)&$\alpha$\\
\hline 
MIN & $1$ & $0.00595$ & $0.55$ \\ 
MED & $4$ & $0.0112$ & $0.70$ \\
MAX & $15$ & $0.0765$ & $0.46$ \\
\hline 
\end{tabular}
\caption{{\footnotesize Values of propagation parameters
widely used in the literature and roughly providing minimal and maximal positron fluxes,
or constitute the best fit to the B/C data.}}
\label{PropParametersPos}
\end{table}
\end{center}

As in the case of antiprotons, a crucial factor in the treatment of positron propagation is the adopted 
propagation model. In this case there are mainly 3 relevant parameters, namely $L$, $K_0$ and $\alpha$; that is, 
the half-thickness of the cylindrical diffusive zone, the diffusion constant and the spectral index, respectively. For their values we use the three models described in  table \ref{PropParametersPos}. 

\subsection{The primary flux}

The resulting positron flux  from DM annihilations can be written as (see \cite{TheorUncertainties} for details)
\begin{equation}
\Phi_{e^+} (E)= \frac{\beta_{e^+}}{4\pi} 
 \frac{\left\langle \sigma v \right\rangle}{2}  \left( \frac{\rho(\vec{x}_\odot)}{m_\chi} \right) ^2
\frac{\tau_E}{E^2}
\int_E^{m_\chi} f(E_s) \tilde{I}(\lambda_D) dE_s\,,
\label{PosFlux}
\end{equation}
where the detection and the production energy are denoted respectively by $E$ and $E_s$. $f(E_s)$ is the production spectrum for positrons, $f(E_s) = \sum_{i} dN_{e^+}^i/dE_s$, with $i$ running over all possible annihilation channels. The diffusion length,  $\lambda_D$, is defined by
\begin{equation}
\lambda_D^2 = 4 K_0 \tau_E \left(\frac{E^{\alpha-1} - E_s^{\alpha-1}}{1-\alpha} \right) .
\end{equation}
The so-called halo function, $\tilde{I}$, contains all the dependence on astrophysical factors. It is given as
\begin{equation}
\tilde{I}(\lambda_D) = \int_{DZ} d^3\vec{x}_s \tilde{G}(\vec{x}_\odot, E \rightarrow \vec{x}_s, E_s)
\left( \frac{\rho(\vec{x}_s)}{\rho(\vec{x}_\odot)} \right) \,.
\end{equation}
The modified Green function $\tilde{G}$ is in its turn defined by 
\begin{equation}
\tilde{G} = \frac{1}{4\pi K_0 \tilde{\tau}} e^{-(z_\odot - z_s)^2/4 K_0 \tilde{\tau}} \tilde{V}\,,
\end{equation}
with $\tilde{V}$ depending on the value of the characteristic parameter $\zeta=\frac{L^2}{K_0 \tilde{\tau}}$. When $\zeta>1$  --when the diffusion time is small-- boundary conditions can be ignored and the propagation equation can be treated as a $1$-$D$ Schroedinger equation.  In that case
\begin{equation}
\tilde{V} = 
\frac{1}{\sqrt{4 \pi K_0 \tilde{\tau}}}
\exp \left(  -\frac{(z-z_s)^2}{4 \pi K_0 \tilde{\tau}} \right)\,. 
\end{equation}
When $\zeta$ is small this approximation no longer holds but  we can express
$\tilde{V}$ as
\begin{equation}
\tilde{V} = \sum_{n = 1}^{\infty} 
\frac{1}{L} \left[ e^{-\lambda_n \tilde{\tau}} \phi_n(z_s)  \phi_n(z) + 
 e^{-\lambda_n' \tilde{\tau}} \phi_n'(z_s)  \phi_n'(z)   \right] 
\end{equation}
where
\begin{align}
\phi_n(z)  &= \sin(k_n(L - |z|))\,,  &k_n =&  \left( n - \frac{1}{2}\right) \frac{\pi}{L}\\ 
\phi_n'(z) &= \sin(k_n'(L - z))\,,   &k_n' =&  n \frac{\pi}{L} \,.
\end{align}

We now have all the necessary ingredients for the computation of the expected positron flux from singlet scalar dark matter annihilation. Next we discuss the positron background.

\subsection{Astrophysical Backgrounds}
In the conventional background model, positrons are produced in the interactions between cosmic-ray nuclei and the interstellar medium. The expected positron  flux on earth can be written as \cite{Strong:2004de}
\begin{equation}
\Phi_{e^+}^{conv}=\frac{4.5E^{0.7}}{1+650E^{2.3}+1500E^{4.2}}\,\mathrm{GeV}^{-1}\mathrm{cm}^{-2}\mathrm{s}^{-1}\mathrm{sr}^{-1}\,
\end{equation}
where $E$ is given in GeV. In contrast to the antiproton case, this conventional background is not compatible with the recent data from PAMELA \cite{PAMELApositrons} and Fermi-LAT \cite{Abdo:2009zk}. Even after taking into account the possible uncertainties due to cosmic ray propagation, the data reveals a clear excess over this background at high energies, $E\gtrsim 10$ GeV. Hence, a new source of high energy positrons is necessary to explain the data.

Two main interpretations of this positron excess have been considered in the literature. One is the annihilation of dark matter particles. Though viable, such models are rather contrived. They require multi-TeV masses, annihilation into leptonic final states, and an annihilation rate several orders of magnitude larger than that expected for a thermal relic \cite{Meade:2009iu}. Moreover, they are tightly constrained by experimental data --gamma rays, radio emission, CMB, BBN \cite{Bertone:2008xr}. Given that, in addition, the singlet model cannot explain the positron excess (at least in the energy region we examine and for reasonable astrophysical assumptions), we will not consider this dark matter interpretation any further.

The other main interpretation for the observed positron excess is that it is generated by astrophysical sources, such as pulsars \cite{Hooper:2008kg}. If that is the case,  such astrophysical positrons constitute an additional background to the positrons from dark matter annihilation. And they  must be taken into account when studying the sensitivity of future experiments to a dark matter positron signal, as pointed out in \cite{Choi:2009qc}. Here, we will closely follow their approach. That is, we use as background the positron flux obtained from the pulsar interpretation of PAMELA and Fermi-LAT data \cite{Grasso:2009ma}--see figure 2 in \cite{Choi:2009qc}. That background is compatible with the absolute positron flux derived in \cite{Balazs:2009wm}.

\subsection{Results}
\begin{figure}[tb!]
\centering
      \includegraphics[width=0.40\textwidth,angle=-90]{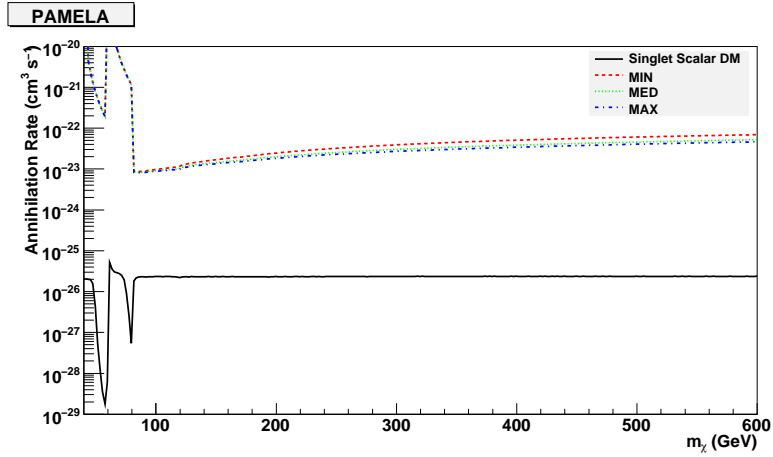}
      \caption{\footnotesize Regions of the parameter space that are  excluded by the  recent positron data from the PAMELA experiment. The area above the lines is excluded for the corresponding propagation model. Notice that no region of the viable parameter space is currently ruled out.}
       \label{fig:pospamela}
\end{figure}

For positrons we follow a procedure analogous to that for antiprotons. We first find the regions that are already excluded by present data and then determine those that will be detectable in AMS-02. When using equation (\ref{chi2}), it must be taken into account that  the geometrical acceptance of AMS-02 for positrons is $420\mathrm{cm^2\, sr}$ \cite{Goy:2006pw}.  

Figure \ref{fig:pospamela} displays the regions that are already excluded for the MIN, MED, and MAX propagation models. The solid (black) line shows, instead, the prediction of the singlet scalar model. Notice that the variation due to the propagation model is very small, certainly much smaller than for antiprotons. From the figure we also see that the exclusion lines  all lie well above the model's prediction. We checked that a similar result is found even in the optimistic scenario for substructure enhancement. At present, henceforth, no region of the viable parameter space is even close to being excluded via the positron signal.

\begin{figure}[tb!]
\centering
      \includegraphics[width=0.40\textwidth,angle=-90]{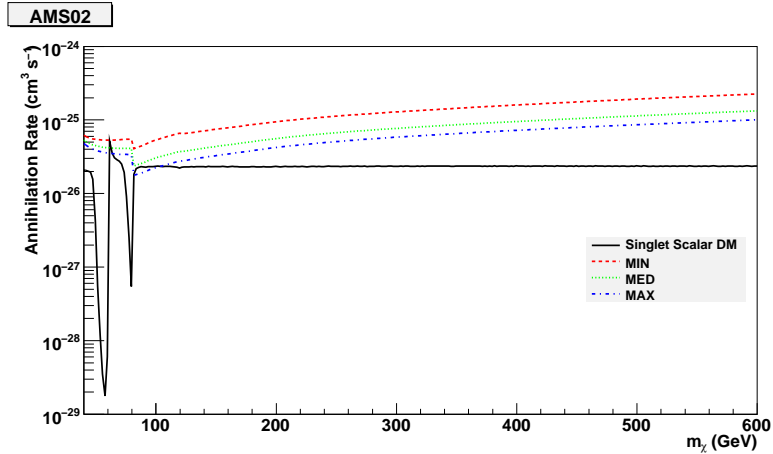}
      \caption{\footnotesize Regions of the parameter space that give a positron signal within the sensitivity of the AMS-02 experiment. The lines corresponding to the MIN, MED and MAX propagation models are shown. They must be compared to the actual prediction (solid line) of the singlet scalar model. }
       \label{fig:posams}
\end{figure}

The AMS-02 detectable regions are shown in figure \ref{fig:posams} for the three propagation models. Again, not much difference is observed between the MIN, MED and MAX lines. Moreover, only a small region around $m_S\sim 200$ GeV is within the AMS-02 reach. Thus, without additional boost factors, singlet masses above $200$ GeV do not produce a detectable positron signal at AMS-02.

\begin{figure}[tbp!]
\centering
	\includegraphics[width=0.40\textwidth,clip=true,angle=-90]{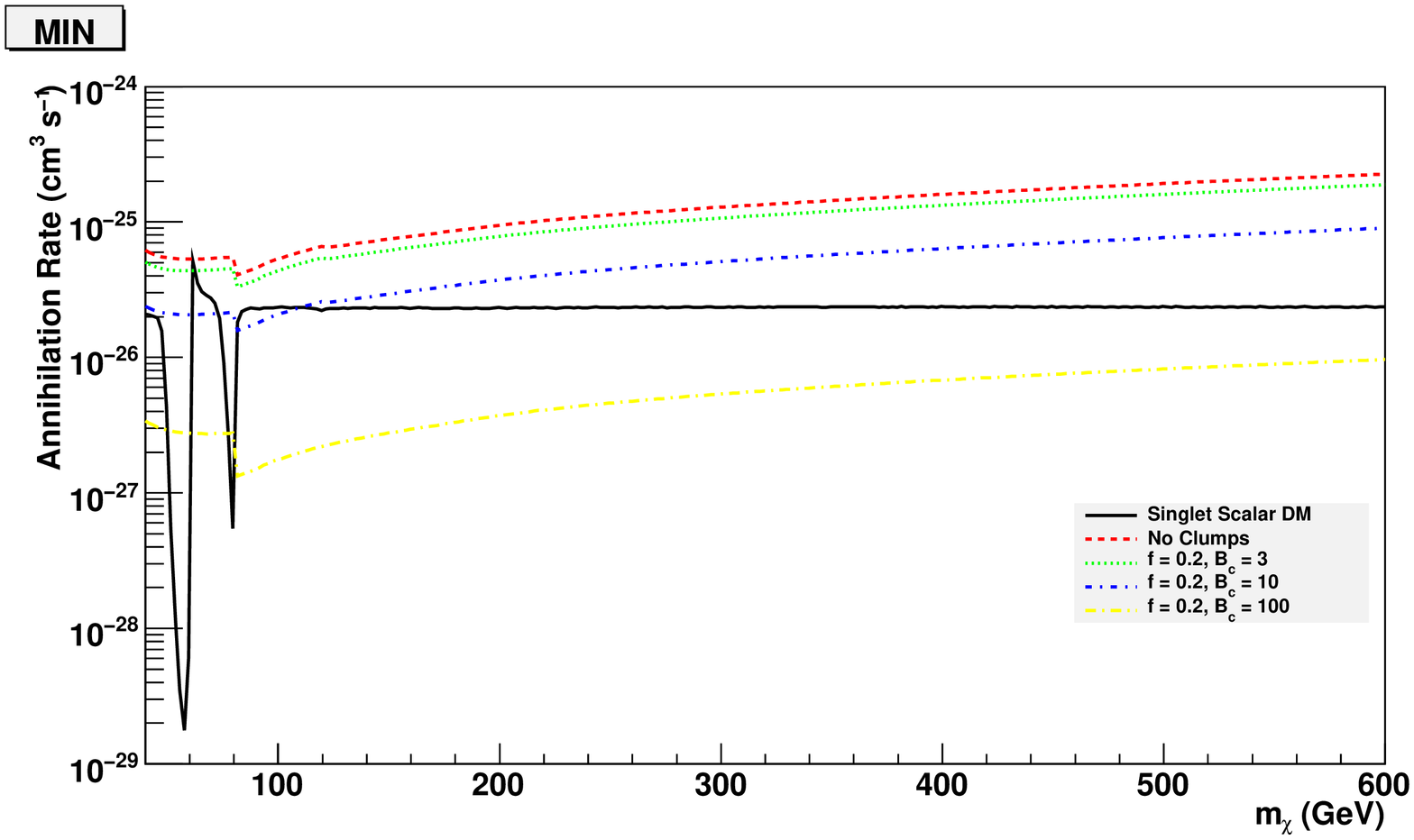}
	\includegraphics[width=0.40\textwidth,clip=true,angle=-90]{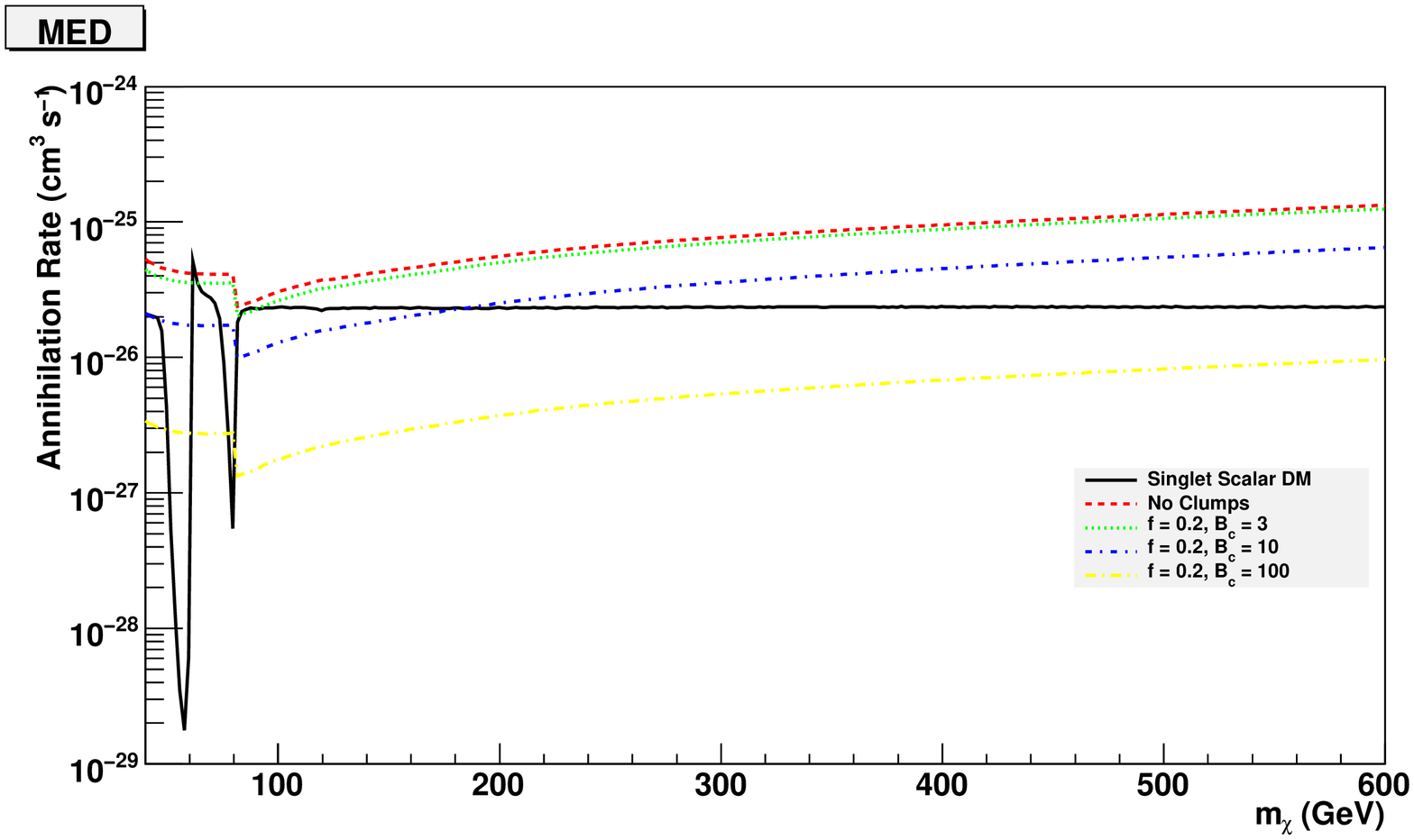}
	\includegraphics[width=0.40\textwidth,clip=true,angle=-90]{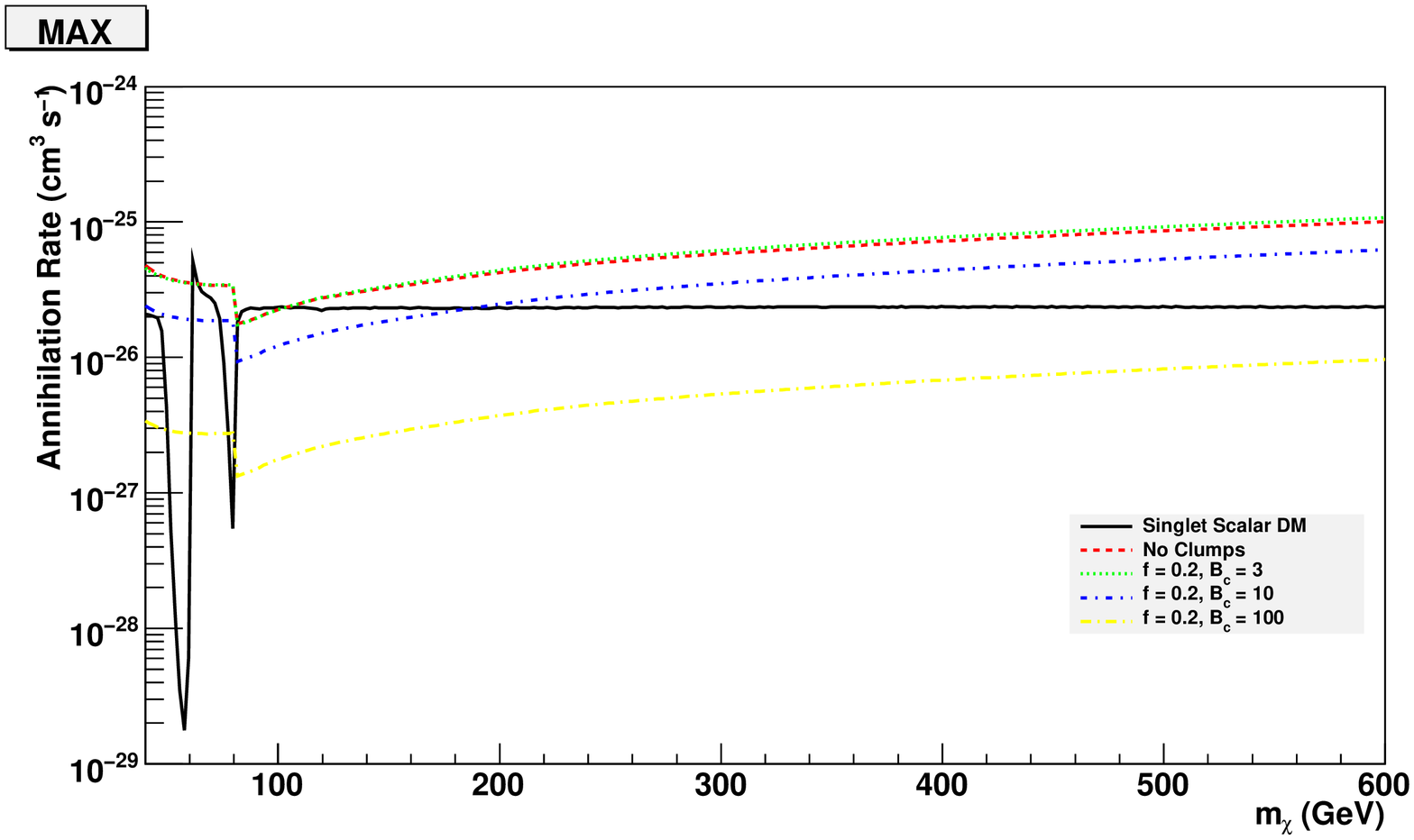}

          \caption{{\footnotesize
Detectable regions for the MIN, MED, and MAX propagation models in the presence of dark matter substructures. The region above the lines is detectable for the corresponding parameters. The solid (black) line shows the prediction of the singlet scalar model. 
}}
\label{fig:ClumpsPosAMS}
\end{figure}

Once the effect of dark matter substructures is included, the detectable region extends to higher singlet masses, as illustrated in figure \ref{fig:ClumpsPosAMS}. For the optimistic scenario, $B_c=100$, the whole mass range becomes detectable independently of the propagation model. For the moderate scenario, $B_c=10$,  the AMS-02 reach extends to singlet masses of $100$ GeV and $200$ GeV for the MIN and MAX models respectively. Thus, a substructure enhancement is required to probe the singlet model beyond $m_S\sim 200$ GeV at AMS-02.

\section{Discussion}
\label{sec:disc}

Our results clearly indicate that in the singlet scalar model the indirect detection of dark matter via antimatter is feasible, particularly through the antiproton channel. The positron channel also gives rise to an observable signal at AMS-02, but only for light singlets.  The antiproton signal, on the other hand, should be visible over most of the viable parameter space.

To identify the singlet scalar model as the correct model of dark matter, one would need to measure $m_h$ and $m_S$, the only two parameters of the model, and to confirm some of the model predictions. Such a task would certainly require input from different experiments: accelerator searches, direct detection experiments, and indirect searches through several channels. Searches at the LHC, for instance, could provide a measurement of the higgs mass, but they are not expected to say much about the singlet mass. It is mainly through direct and indirect detection searches that the singlet mass could be determined.

In \cite{Yaguna:2008hd}, it was shown that the singlet direct detection rate is within the reach of current and  future experiments and that the gamma ray signal from singlet dark matter annihilation is likely to be observed by the Fermi satellite. In this paper, we have found that also the antiproton signal is observable at AMS-02.  By combining the signals from these three experiments the singlet mass, $m_S$, might be determined. Once $m_h$ and $m_S$ are known, the singlet model of dark matter is completely specified, at least within the standard cosmological scenario.   

\section{Conclusions}
\label{sec:conc}
The singlet scalar model provides a simple and predictive scenario to account for the dark matter. In this paper, we studied in detail the antiproton and positron  signals expected from the annihilation of singlet scalar dark matter. Particular attention was paid to the role of the propagation model and to a possible enhancement in the annihilation rate due to dark matter substructures in the Galaxy. We used the recent measurements, from PAMELA and Fermi-LAT, of the  positron and antiproton flux  to find the regions in the plane ($m_S,\sigmav$) that are already excluded by the data. For antiprotons,  an optimistic enhancement, $B_c=100$, due to dark matter clumps is essentially excluded for both the  MED and the MAX propagation models in the entire mass range we considered. For positrons, on the contrary, no region is excluded independently of the propagation model or the possible effect of substructures. We also analyzed the perspectives for detecting antimatter signals in future experiments.  In AMS-02, the positron signal, in the absence of additional boosts, was found to be detectable but only for light singlets. Antiprotons, on the other hand, offer excellent perspectives to be detected at AMS-02. Even without boosts, the whole mass range is within the sensitivity of AMS-02 for the MED and the MAX propagation models. For the MIN model, a moderate enhancement from substructures, $B_c=10$, would allow to probe the mass range below $200 \gev$.  Antiprotons, therefore, seem to be a promising way to constrain or detect the singlet scalar model of dark matter.

\section*{Acknowledgments}
Carlos E. Yaguna  is supported by the \emph{Juan de la Cierva} program of the Ministerio de Educacion y Ciencia of Spain. He receives additional support from  Proyecto Nacional FPA2009-08958,  the ENTApP Network of the ILIAS project RII3-CT-2004-506222, and the Universet Network MRTN-CT-2006-035863. The work of Andreas Goudelis is supported in part by the E.C. Research Training Networks under contract MRTN-CT-2006-035505.

\end{document}